\documentclass[usegraphicx,usenatbib]{mn2e}

\usepackage{times}
\usepackage{verbatim}

\newcommand{\plotone}[1]{\includegraphics[width=\columnwidth]{#1}}
\newcommand{\apj}{ApJ}           
\newcommand{\mnras}{MNRAS}       
\newcommand{\aap}{A\&A}

\newcommand{\apjs}{ApJS}
\newcommand{\aj}{AJ}
\newcommand{\pasp}{PASP}

\title[SDSS spectra in 4DE1: Radio-Loud/Radio-Quiet Dichotomy]
{New Insights on the QSO Radio-Loud/Radio-Quiet Dichotomy: \\
      SDSS Spectra in the Context of the 4D Eigenvector1 Parameter Space}

\author[Zamfir et al.]{S. Zamfir$^1$\thanks{E-mail:
zamfi001@bama.ua.edu}, J.~W.\ Sulentic$^1$ and
P.\ Marziani$^2$\\
$^1$Department of Physics and Astronomy, University of
Alabama, Box 970324, Tuscaloosa, AL 35487, USA\\
$^2$INAF-Osservatorio Astronomico di Padova, Vicolo
dell'Osservatorio 5, I-35122 Padova, Italy\\
}

\pagerange{\pageref{firstpage}--\pageref{lastpage}} \pubyear{2007}

\begin{document}
\label{firstpage} \maketitle

\begin{abstract}

We search for  a dichotomy/bimodality between Radio Loud (RL) and
Radio Quiet (RQ) Type 1 Active Galactic Nuclei (AGN). We examine
several samples of SDSS QSOs  with high S/N optical spectra and
matching FIRST/NVSS radio observations. We use the radio data to
identify the weakest RL sources with FRII structure to define a
RL/RQ boundary which corresponds to log L$_{1.4GHz}$=31.6 ergs
s$^{-1}$ Hz$^{-1}$. We measure properties of broad line H$\beta$ and
FeII emission to define the optical plane of a 4DE1 spectroscopic
diagnostic space. The RL quasars occupy a much more restricted
domain in this optical plane compared to the RQ sources, which a 2D
Kolmogorov-Smirnov test finds to be highly significant. This tells
us that the range of BLR kinematics and structure for RL sources is
more restricted than for the RQ QSOs, which supports the notion of
dichotomy. FRII and CD RL sources also show significant 4DE1 domain
differences that likely reflect differences in line of sight
orientation (inclined vs. face-on respectively) for these two
classes. The possibility of a distinct Radio Intermediate (RI)
population between RQ and RL source is disfavored because a 4DE1
diagnostic space comparison shows no difference between RI and RQ
sources. We show that searches for dichotomy in radio vs. bolometric
luminosity diagrams will yield ambiguous results mainly because in a
reasonably complete sample the radio brightest RQ sources will be
numerous enough to blur the gap between RQ and RL sources. Within
resolution constraints of NVSS and FIRST we find no FRI sources
among the broad line quasar population.

\end{abstract}

\section{Introduction}

A much debated problem in AGN studies involves the possibility of a
real physical dichotomy between radio-loud (RL) and radio-quiet (RQ)
QSOs. The low fraction of RL sources - on average $\sim$ 5-25$\%$
\citep[e.g.][]{Kellermann89, Padovani93, Kellermann94, Jiang07}
depending on the adopted definition of radio-loudness - and its
dependence on redshift and optical luminosity \citep[e.g.][]{PML86,
Miller90, Visnovsky92, Padovani93, Hooper95, Goldschmidt99, Jiang07}
add to the difficulty of defining statistically meaningful samples
of QSOs with which to identify potentially bimodal properties.

Another complication is introduced by the fact that some good
fraction of RQ sources share common properties with the RL quasars;
for example: a) about 30-40\% of RQ QSOs are spectroscopically
similar to RL \citep[e.g.][and present study]{Sulentic00a} and b)
both QSO types are capable of producing radio jets. RQ jets
typically extend over scales of a few parsecs up to kiloparsecs
\citep[e.g.][]{Blundell98, Kukula98, Ulvestad05,
Leipski06}\footnote{However \citet{Ulvestad05} report deep lower
frequency VLBI observations of several RQ objects studied by
\citet{Blundell98} and do not confirm the presence of jet-related
structure.} and RL much larger scales with higher radio power
\citep[e.g.][]{RS91, Miller93}. Potential bimodal properties might
be hiding behind such similarities.

The very definition of radio-loudness is rather ``loose'' with
continued disagreement over the empirical RL/RQ boundary. Over the
last few decades a couple of possible boundary criteria have been
proposed based on: Criterion 1 - radio power \citep{Miller90} and
Criterion 2 - radio/optical flux density ratio. Criterion 2 involves
the much used Kellermann factor R$_{K}$ (radio flux density at 6cm
normalized to B-band flux density). \citet{Kellermann89} suggested
R$_{K}$ = 10 for the RL-RQ boundary in the Palomar-Green (PG) sample
of QSOs \citep{SG83, GSL86, BG92} and many studies have adopted this
value. Others have suggested that different nominal R$_{K}$ limits
for radio steep- and flat-spectrum sources would eliminate the
confusion introduced by a fixed value of 10 \citep{Falcke96a}.
\citet{Sikora07} propose another kind of quantitative distinction
for RL AGNs based on R$_{K}$ as a function of Eddington ratio (see
their section 4.1). This definition might be relevant if one
includes FRI sources and LINERs that do not show broad lines.

There are several different surrogate definitions of R$_{K}$ in
literature involving radio measures at various frequencies and
optical (B-band or i-band), UV or even X-ray measures (e.g.
\citealt{Kellermann89, Stocke92, Ivezic02, TW03, Jester05a, Wang06,
Jiang07}), which obviously complicates the comparison of different
studies.

It is still unclear whether one of the two criteria is more
physically significant. Studies like \citet{Miller90} promote the
radio power as a more fundamental discriminator, while others argue
in favor of the second criterion, which relates the radio properties
to other regimes of energy output. Moreover, while some galactic
nuclei qualify as radio-loud based on one boundary criterion they
fail to do so when using the other. A good example \citep{HP01}
involves a sample of bright Seyfert nuclei where as many as
$\sim$60\% of the sources are RL using R$_{K}$ $>$ 10 (adopting a
``nuclear'' radio-optical ratio), but only one would satisfy the
condition of L(6cm) $>$ 10$^{25}$ W Hz$^{-1}$ sr$^{-1}$
\citep{Miller90}. The conclusions in \citet{HP01} are provocative in
terms of both the radio loud fraction and the fact that many of
their so-called ``radio-loud'' would be hosted by spiral galaxies, a
rather different result compared to more luminous samples analyzed
in studies like e.g. \citet{Taylor96}, \citet{McLure99},
\citet{Dunlop03}. \citet{Laor03} explains (based on the results of
\citealt{Xu99}) that R$_{K}$ is Luminosity-dependent and one should
rather use R$_{K}$$\propto$ L$^{-0.5}$ to separate RLs from RQs. The
R$_{K}$=10 suggested by \citet{Kellermann89} for luminous samples
(M$_{B}$$\sim$-26) is not a valid choice for RL boundary for low
luminosity samples \citep[e.g.][]{HP01}. Nonetheless, R$_{K}$
retains its heuristic value because it offers a scaling relation
between nonthermal and thermal mechanisms at work in AGN. After all,
for theoretical models of accretion disk it is preferable to use
dimensionless quantities like the Eddington-scaled luminosity and
accretion rate (although there may be different scaling relations
for jets, disk, corona luminosity with accretion rate and black hole
mass, e.g. \citealt{Kording06a}).

Thus, we face the problem of labeling objects differently depending
on the adopted definition of boundary criterion, which complicates
the integration of different results into a more general picture. A
further problem involves the combination of radio and optical flux
measures (the latter being susceptible to internal extinction) that
can introduce serious selection effects and biases at different
redshifts thus making R$_{K}$ a problematic radio-loudness indicator
for statistical purposes. These aspects reenforce the necessity of
alternative approaches toward a consistent definition of
\textit{radio-loud}. We need a more standardized definition.

Different studies over the last decade report contradictory results
regarding the question of a bimodal distribution of QSOs in terms of
radio-loudness. Recent SDSS-based studies \citep[e.g.][]{Ivezic02,
White07} defend the reality of bimodality for QSO distribution using
histograms of radio/optical(UV) ratios. The latter study shows a
significant dip at R$_{K}$$\sim$30-40 (see their Figure 15). They
employ image stacking to lower the detection limit of FIRST to
nano-Jy levels and their final sample includes over 41000 sources. A
bimodal distribution is found by \citet{Liu06} in terms of R$_{K}$
corrected for orientation (although with a heterogeneous sample).
Their Figure 9 shows a cutoff in the RL population in the range
logR$_{K}$ = 1.5-2.0. There are at the same time many studies that
question the reality of bimodality \citep{Falcke96a, White00,
Brotherton01, Lacy01, Cirasuolo03a, Cirasuolo03b}.
\citet{Falcke96a}, \citet{Lacy01}, \citet{Brotherton01} propose a
population of Radio Intermediate sources (RI) that might bridge the
gap between RL and RQ. Clearly, the RL/RQ problem is far from
resolved reflected in the lack of consensus on how to consistently
define a radio-loud sample or prove the existence of a physical
dichotomy.

Even if the distribution of radio-loudness measures for a sample of
RL and RQ sources may not exhibit bimodality, we showed that they
may represent two distinct classes of AGN, based on spectroscopic
measures \citep{Sulentic03}. The present study attempts to provide
more robust empirical support to this alternative approach. This
paper also shows that the picture of ``dichotomy'' is significantly
distorted by mixing bright and faint QSO samples when we are
flux-limited in both optical and radio regimes.

We recently considered \citep{Sulentic03} a third RL/RQ boundary
criterion (Criterion 3) based on the classical radio morphology.
Double-lobe FRII morphology \citep{FR74} is the most common type
observed in broad line emitting RL quasars. FRI morphology is very
rare among broad-line AGN, e.g. 3C120, E1821+643 \citep{BR01} and
SDSS J104022.79+444936.7 \citep{Heywood07}. The latter reference
suggests that FRI morphology may become more common among broad-line
quasars beyond z$\sim$1.0. In a simple orientation unification
scenario \citep{UP95} core-dominated (CD) RL sources are interpreted
as FRII sources viewed with radio jet axis aligned close to our line
of sight. If this scenario is valid then the CD counterparts of any
FRII population will be on average more radio luminous due to
relativistic boosting effects \citep[e.g.][]{OB82, Scheuer87,
Barthel89, JW99}. We adopted the weakest (assumed unboosted) FRII
sources in our sample to define the lower boundary of the RL
phenomenon. This also allowed us to redefine the boundary in terms
of Criterion 1 (log$P_{6cm}$$\sim$32.0 erg s$^{-1}$ $Hz^{-1}$) and
Criterion 2 (R$_{K}$ $\sim$ 70).

In the present study we reiterate the idea that a robust definition
of a radio-loud quasar can be formulated only if radio-morphology is
taken into account. More specifically the FRIIs should be considered
the parent population of RL quasars \citep[e.g.][]{OB82, Scheuer87,
Barthel89, Taylor96, JW99}. This is also supported by the recent
confirmation that the shape of the observed luminosity function of
FRII radio galaxies is the same as the intrinsic luminosity function
of RL quasars \citep{Liu07}. In other words, the radio weakest FRII
structures should dictate the empirical boundary between RL quasars
and the rest of the QSO population. We therefore defined a sample of
RL quasars using the radio luminosity coupled with the radio
morphology in order to avoid the perviously discussed problems
associated with R$_{K}$ (see also \citealt{WK99}).

We have been exploring a 4D parameter space
\citep[4DE1;][]{Sulentic00a, Sulentic00b, Marziani01, Marziani03a,
Marziani03b, Sulentic07} that serves as a spectroscopic
unifier/discriminator for all broad emission line AGNs (Type 1). Our
principal parameters involve measures of: 1) full width at half
maximum of broad H$\beta$ (FWHM H$\beta$), 2) equivalent width ratio
of optical FeII ($\lambda$4570\AA\ blend) and broad H$\beta$,
R$_{FeII}$=W(FeII $\lambda$4570\AA)/W(H$\beta$), 3) the soft X-ray
photon index ($\Gamma_{soft}$) and 4) CIV$\lambda$1549\AA\ broad
line profile velocity displacement at half maximum, c(1/2). The
``Introduction'' of \citet{Sulentic07} explains how this parameter
space evolved from various pioneering works. One enormous advantage
of this parameter space formulation is its weak or absent dependence
on source luminosity \citep[][]{Sulentic00a, Sulentic04}. Armed with
our improved definition of the lower boundary for RL activity in
quasars we compared their 4DE1 properties with RQ sources. We found
that most RL sources show a much restricted domain space occupation
within the optical parameter plane of 4DE1 (FWHM H$\beta$ vs.
R$_{FeII}$) compared to the RQ majority \citep{Sulentic03}.

Although the results presented in 2003 provided compelling  support
for RL-RQ bimodality, the adopted sample was rather heterogeneous,
incomplete and included many sources with measures derived from
marginal S/N spectra. With the advent of the SDSS database it
becomes possible to select a large and much more complete ($\sim$
90$\%$)\footnote{As defined within the SDSS project the completeness
was estimated in two ways: by checking how many of the previously
known QSOs are recovered and evaluating the output of target
selection for simulated quasars, see \citet{Richards02} and
\citet{Vandenberk05}} sample of AGN with uniformly high resolution
and S/N optical spectra. A further advantage involves the larger
wavelength interval 3800-9200{\AA} sampled by SDSS. This is
complemented by uniform radio survey data from FIRST\footnote{Faint
Images of the Radio Sky at Twenty-centimeters (FIRST)-
http://sundog.stsci.edu/; see also \citet{Becker95}} designed to
match the SDSS sky coverage and NVSS survey\footnote{NRAO VLA Sky
Survey (NVSS) - http://www.cv.nrao.edu/nvss/; see also
\citet{Condon98}}.

The value of studying the radio-loud phenomenon within the 4DE1
context is at least twofold: 1) it compares RL and RQ sources in a
parameter space defined by measures with no obvious dependence on
the radio properties \citep{Marziani03b} and 2) it allows us to make
predictions about the probability of radio loudness for any
population of QSOs with specific optical (or UV) spectroscopic
properties.

The paper is organized as follows: \S~2 presents the sample
selection and the RL definition based on radio morphology and radio
luminosity. \S~3 presents the 4DE1 optical measures. \S~4 includes
differences in RQ, RI and RL source occupation within 4DE1. \S~5
offers a discussion on the RL/RQ dichotomy based on L-dependent
diagrams. In \S~6 and \S~7 we discuss the fraction of RL quasars and
the probability of radio-loudness within the 4DE1 optical plane. The
last two sections are dedicated to discussions and conclusions.
Throughout this paper we use H$_{o}$ = 70 km s$^{-1}$ Mpc$^{-1}$,
$\Omega_{M}=0.3$ and $\Omega_{\Lambda}=0.7$.

\section{Defining a Population of Radio-Loud Quasars}

We consider any AGN that shows broad (Balmer) emission lines as
``QSO'' regardless of its intrinsic luminosity (or absolute
magnitude). This is why we generated our own sample of SDSS QSOs
(from Data Release 5 of SDSS; \citealt{Adelman07}) instead of
extracting it from the vetted catalog of \citet{Schneider07} that is
limited to sources with absolute magnitude M$_{i}$ $\leq$ -22.0.
Sample size is driven by the following goals: i) a sample of high
quality spectra suitable for 4DE1 spectroscopic analysis, ii) as
complete as possible sample of RL quasars, iii) a large enough
sample of RQ quasars with reliable 4DE1 measures so that we can
define the RQ zone of occupation, iv) a representative sample of
so-called RI sources and v) source-by-source evaluation to avoid the
pitfalls (e.g. radio/optical misidentifications, misclassifications)
of automated processing. Our approach is based on careful
examination of each optical spectrum in order to confirm the
presence of broad lines. All FIRST/NVSS (20cm/1.4GHz) radio maps
were visually examined in order to evaluate radio morphology,
resolve ambiguous cases and obtain the correct integrated (total)
radio flux density.

OPTICAL SELECTION: We restricted source selection to z $\leq$ 0.7 so
that H$\beta$ and adjacent spectral regions of interest (used to
define the underlying continuum) would be accessible. We selected
our sample in several steps. Step 1 selected all SDSS DR5 quasars
with psf g$<$ 17.0 (n=333 QSOs with 34 RL). This selection was
motivated by the need for high S/N spectra from which 4DE1
parameters could be reliably measured. Step 2 extended this limit to
psf g=17.5 (n=806 QSOs with 76 RL). This extension was motivated by
the desire to increase the RL sample. Our first two steps are based
on a rather blue filter, close to BQS \citep{SG83, GSL86, BG92}.
Although \citet{Jester05a} find no radio-related incompleteness for
BQS-like selected samples, they point out that the apparently large
fraction of RL in BQS survey is related to its rather blue filter
(B-band) selection (see \S~6). Having this in mind, we considered
also a step 3 aimed to define a RL sample considering all QSOs
brighter than psf i=17.5 (n=1656 QSOs with 91 RL). In Table 1 we
explain which objects have been considered for the spectroscopic
analysis.

\begin{table*}
\caption{Samples with z $<$ 0.7 selected for spectral analysis in
the Context of 4DE1.} \tabcolsep=3pt
\begin{tabular}{lcccc}
\hline
Apparent Magnitude & N$_{total}$ & N$_{RL}$ & N$_{radio-detected}$ & Considered for Spectral Analysis?\\
\hline\hline

psf g $<$ 17.0 & 333 & 34 & 136 & all\\

17.0 $\leq$ psf g $<$ 17.5 & 473 & 42 & 122 & all radio detected (either FIRST or NVSS)\\

psf i $<$ 17.5 AND psf g $\geq$ 17.5 & 850 & 19 & 206 & only if
L$_{1.4GHz}$ $\geq$ 31.0 erg s$^{-1}$ Hz$^{-1}$ (all RI and RL)\\

\hline
\end{tabular}
\begin{minipage}{17.8cm}
Notes: Every QSO in the samples selected based on g-filter was
examined in FIRST/NVSS radio maps.

The radio properties for the sample of QSOs selected by the third
set of criteria (based on i-filter) were determined in several
steps: 1) we searched FIRST within 15$\arcsec$ from the optical
position, then 2) all sources without a detection (or not in area
covered by FIRST) were searched in NVSS around 80$\arcsec$ and
finally 3) all sources that are radio-detected in either survey are
examined in detail in order to get the total flux density from all
components.

\end{minipage}

\end{table*}

We offer in Figure 1 the sources redshift distributions resulting
from the two selections (based on psf g $<$ 17.5 and based on psf i
$<$ 17.5). Selection using the g-band magnitude limit yields a
sample with much more uniform redshift distribution. Not
surprisingly i-band selection (much more complete) favors redder
local QSOs (50\% have z $<$ 0.15; 60\% have z $<$ 0.20 and 70\% have
z$<$ 0.25), more strongly affected by host galaxy contamination
and/or by the presence of H$\alpha$ line inside i-band. There is
obviously a large overlap between the samples. The census presented
here also includes the objects that show broad lines narrower than
1000 km s$^{-1}$ and are labeled ``Galaxy'' by the SDSS
spectroscopic pipeline (see the appendix and see also Table 2). Such
sources can be safely included under the generic umbrella of
``QSO''. However, these objects were not included in our
spectroscopically-processed sample for the following reasons: 1)
limitation of our template used to extract the FeII lines from our
spectra (same as used in \citealt{BG92}) or 2) very red continua,
extreme Balmer decrement, significant galaxy contamination and
dramatically different H$\alpha$ and H$\beta$ broad lines. We also
included all QSO spectra that have been assigned only a ``Science
Primary'' index of 0\footnote{``The SDSS SCIENCEPRIMARY flag
indicates whether a spectrum was taken as a normal science spectrum
(SCIENCEPRIMARY = 1) or for another purpose (SCIENCEPRIMARY= 0). The
latter category contains quality assurance and calibration spectra,
or spectra of objects located outside of the nominal survey area.''
\citep{Schneider07}}.

\begin{figure}
\plotone{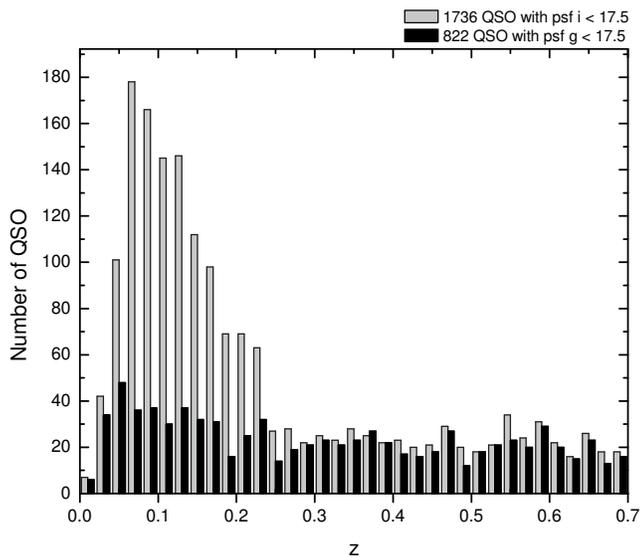} \caption{The distribution of QSOs apparently
brighter than psf g = 17.5 (black) and brighter than psf i = 17.5
(light grey).}
\end{figure}

RADIO SELECTION: We used FIRST combined with NVSS to evaluate the
integrated radio emission and source structure. For FIRST survey the
typical rms fluctuations are 0.15 mJy, and the resolution is 5". For
NVSS survey the rms brightness fluctuations are 0.45 mJy
beam$^{-1}$, with a 45 \arcsec\ resolution (see footnotes 3 and 4 of
this paper for the sources of this technical details). Both radio
maps were compared to avoid missing extended sources that might have
been attenuated with FIRST. The 45\arcsec\ beam of NVSS yields
sensitivity to more extended structure and provides radio data for a
few sources not observed by FIRST. We also wanted to clarify the
nature of any significant discrepancies between the two surveys for
specific sources.

\begin{table*}
\caption{Additional samples with z $<$ 0.7 used in this study, but
not measured spectroscopically. We require that they all show
bona-fide Type 1 QSO spectra.} \tabcolsep=3pt
\begin{tabular}{lcccc}
\hline
Apparent Magnitude & Category/Type & N$_{total}$ & N$_{radio-detected}$ & N$_{RL}$\\
\hline\hline

psf g $<$ 17.5 & ``Galaxy''-labeled by SDSS & 16 & 12 & 0\\
psf i $<$ 17.5 AND psf g $\geq$ 17.5 & ``Galaxy''-labeled by SDSS & 81 & 33 & 1\\
psf g $\geq$ 17.5 AND psf i $\geq$ 17.5 & double-lobed (FRII) from \citet{deVries06} & 67 & 67 & 67 \\
19.0 $\leq$ psf g $<$ 19.5 & ``QSO''-labeled by SDSS & 4800+ & 134 & 47\\
19.0 $\leq$ psf i $<$ 19.5 & ``QSO''-labeled by SDSS & 3800+ & 80 & 31\\

\hline
\end{tabular}
\begin{minipage}{17.8cm}
Notes: For the last two samples listed in the table one can notice
the very large N$_{total}$ and a very small N$_{radio-detected}$. We
required a SDSS/FIRST optical match within one arcsec. We are not
concerned about completeness of these subsamples used in \S~5 and
\S~6.\\
Our sample already included all sources from \citet{deVries06} that
have psf g $<$ 17.5 OR psf i $<$ 17.5.
\end{minipage}
\end{table*}

We identified all bona fide FRII structures associated with our
sample of QSOs. Two sources (SDSSJ075407.96+431610.6 and
SDSSJ080131.97+473616.0) show bright cores with detached and
apparently unrelated satellite sources using FIRST. NVSS shows that
they are FRII with very large (Mpc-scale) radio FRII structure that
reveals the satellite sources as associated hotspots. SDSS
J013521.67-004402.0 is an excellent example of false FRII and false
RL. At the same time not all detected double-lobed sources in a
sample can be unambiguously classified as FRII. However all sources
with: 1) low enough redshift , 2) a broad line spectrum (Type 1 AGN)
and 3) adequate radio resolution show FRII (or hybrid e.g. HYMOR -
see \citealt{GW00}) morphology. We therefore assume that all double
lobed RL sources in our sample are FRII. All sources with FRII
structure are assumed to be RL, while the other sources (with core
or core-jet radio morphology) are considered RL only if they have
the radio power L$_{1.4GHz}$ above the threshold set by the weakest
FRII.

Comparison of NVSS and FIRST flux densities for each RL source
reveals: a) NVSS measures are larger than corresponding FIRST values
for virtually all FRII sources and b) most CD RL sources show
agreement between FIRST and NVSS measures with a scatter of
$\sim$$\pm$20 mJy, most likely due to variability \citep[see
also][]{Wang06}. In a few extreme cases we see differences of up to
$\sim$$\pm$300 mJy. There is no evidence for a significant
population of CD RL sources with excess NVSS flux density that might
be due to extended structure not seen by FIRST.

We found n=48 FRIIs brighter than psf g=17.5 OR psf i=17.5. NVSS
flux density measures were always preferred over FIRST (for all
sources, not only for FRIIs). The small size of the FRII sample
motivated us to add double-lobe quasars in our redshift range that
were identified by \citet{deVries06} based on DR3. Careful
reexamination of optical spectra and radio maps for this addition
caused us to eliminate several objects (e.g. no broad emission
lines, no FRII morphology, no radio detection). The DR3-based
sample, which added n=67 FRIIs, was not optically constrained, and
most of the quasars in that subsample showed apparent magnitudes
fainter than our g or i-band limits. Their sample is not meant as an
exhaustive list, but the search algorithm had a reported accuracy of
$\sim$98\% for identifying FRII sources.

\begin{figure}
\plotone{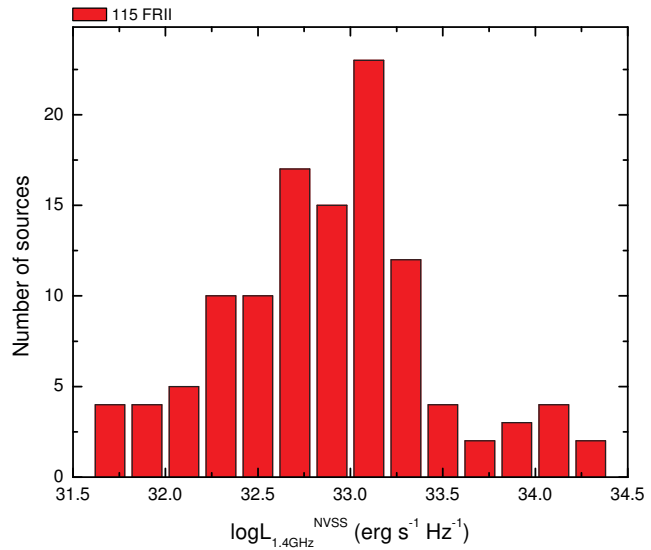} \caption{The distribution of logL$_{1.4GHz}$
radio luminosity - calculated based on NVSS integrated flux density
- for all FRII quasars identified in our sample plus all FRII
sources with z $\leq$ 0.7 from \citet{deVries06}.}
\end{figure}

Figure 2 shows the radio luminosity distribution for the 48+67=115
FRII sources. The weakest bone fide FRII/quasar found shows
logL$_{1.4GHz}$ = 31.6 (erg s$^{-1}$ Hz$^{-1}$). Thus, this becomes
our radio-luminosity defined boundary between RL and RQ QSOs
(Criterion 1). The FRII sources in this sample span three orders of
magnitude in radio luminosity (median logL$_{1.4GHz}$ = 32.9 erg
s$^{-1}$ Hz$^{-1}$), although they become relatively rare above
logL$_{1.4GHz}$$\sim$ 33.5 erg s$^{-1}$ Hz$^{-1}$. Figure 2 shows a
continuous distribution of radio powers, with a clear decline in the
number of sources toward our RL/RQ nominal boundary. The shape of
the distribution suggests that there could be only a very small
number of FRIIs weaker than our weakest bona-fide FRII quasar. It is
beyond the scope of the paper to fully examine the true nature of
the functional form describing the FRII radio-luminosity
distribution. However, we attempted a comparison with the dual
population model for the radio-luminosity function proposed by
\citet{Willott01} for steep spectrum radio sources. We extrapolated
the 1.4GHz measures to 151 MHz measures (to allow for common grounds
with that study) two ways: 1) assuming a spectral slope
$\alpha_{\nu}$ of 0.5 and 2) using the empirical scaling relation
suggested by \citet{Kording08} (their equation (5)). Either way, we
do not see a decline on the lower luminosity side as abrupt as
presented in Figure 3 of \citet{Willott01}. Secondly, the
L$_{151MHz}$ distribution that we get peaks about one decade fainter
($\sim$ 25.5 W Hz$^{-1}$ sr$^{-1}$) than their model suggests for
high luminosity population alone (although their low-luminosity
population could also include FRII sources). While not complete, the
sample we explore is large enough to be assumed representative of
the FRII RL phenomenon in broad line emitting quasars within z =
0.7.

A recent study \citep{Lu07} reported eleven extended SDSS radio
quasars weaker than logL$_{1.4GHz}$$\sim$31.5 erg s$^{-1}$ Hz$^{-1}$
(see their Figure 2). The authors kindly provided us the list of
those sources and we analyzed them one-by-one. Two of them are
included among our RI and are also listed in Table 3 (SDSS
J110717.77+080438.2 and SDSS J120014.08-004638.7). One other source
is also in our sample as RQ with logL$_{1.4GHz}$ = 30.7 erg s$^{-1}$
Hz$^{-1}$ (SDSS J162607.24+335915.2). All other sources are
unambiguously RQ, in some cases offset a few arcseconds from the
optical quasar and were assumed in \citet{Lu07} to be associated
with the active nucleus.

The goal of our work was to search for a dichotomy or gap in radio
properties between RQ and RL sources. Of course, sources are found
with radio intermediate properties \citep[e.g.][]{Falcke96a,
Sulentic03}. Another goal of this paper was to isolate a population
of these radio-intermediate (RI) sources and try to determine if
they form a unique (special) class of quasars. Unlike the non-RL/RL
boundary we have no clear empirical basis for defining a RQ/RI
boundary because all RI show CD radio morphology. The best that we
can do is to isolate a region that is most clearly RI. Considering
that: a) at our sample redshift limit (z=0.7) the minimum detectable
radio luminosity (within FIRST/NVSS) is logL$_{1.4GHz}$ $\simeq$
31.0 erg s$^{-1}$ Hz$^{-1}$ and b) the radio-FIR (Far Infrared)
correlation spans over five decades in luminosity and extends up to
logL$_{1.4GHz}$ $\simeq$ 31.0 erg s$^{-1}$ Hz$^{-1}$
\citep[e.g.][]{Condon92, Yun01}, we see that value as a reasonable
boundary between star-formation and AGN-dominated radio activity
(\citealt{SA91}, \citealt{Kukula98} and \citealt{Haas03} show that
RQ QSOs follow the radio-FIR correlation). We therefore define RI
sources as those with logL$_{1.4GHz}$ in the interval 31.0-31.6 erg
s$^{-1}$ Hz$^{-1}$. This is a much more restricted RI definition
than the one given in \citet{Falcke96a}.

In brief, a source is considered RL if its L$_{1.4GHz}$ radio power
is at least 31.6 erg s$^{-1}$ Hz$^{-1}$. A sources is considered RI
if its L$_{1.4GHz}$ radio power is at least 31.0 erg s$^{-1}$
Hz$^{-1}$, but less than 31.6 erg s$^{-1}$ Hz$^{-1}$. All other
sources are RQ.

Summarizing, we considered the whole sample obtained from the
combination of psf g $<$ 17.5, psf i $<$ 17.5 with an ``OR'' logical
operator. The total number of sources was N=1770 (n=95 RL). The
sample adopted for spectroscopic evaluation (see Table 1) is
constructed as follows:
\begin{enumerate}
\item all RL and RI QSOs that are brighter than either psf g = 17.5
or psf i = 17.5;
\item all RQ QSOs (radio-detected or undetected) that are brighter
than psf g = 17.0 and
\item all RQ QSOs that are radio detected and show psf g in the range
17.0 - 17.5.
\end{enumerate}

We visually examined the SDSS spectrum for every source and rejected
objects without emission lines (e.g. SDSS J075445.67+482350.7),
objects without broad lines (e.g. SDSS J103900.37+414008.7, SDSS
J104451.72+063548.6) or with bad pixels (e.g. SDSS
J113109.49+311405.5, SDSS J145638.81+442755.2, SDSS
J220103.13-005300.2) that prevented reliable line measures in the
region of interest. We rejected one supernova: SDSS
J113323.97+550415.8 - as reported by \citealt{Zhou06}. Two FRII
quasars (SDSS J092414.70+030900.8 and SDSS J123915.39+531414.6)
showed serious host galaxy contamination (psf g - psf i $>$ 1.0 in
both cases), the broad component of H$\beta$ could not be measured.
We also excluded from our analysis objects with W(H$\beta$) $\leq$
20{\AA}, which can be sources in a high continuum phase (e.g. SDSS
J150324.77+475829.6) and/or very red continua with extreme Balmer
decrement, where H$\alpha$ is completely different from H$\beta$
(e.g. SDSS J004508.65+152542.0).

We should emphasize that our spectroscopic analysis does not include
rare objects with extreme R$_{FeII}$ $>$ 2.0 values. Sources with
extremely strong FeII tend to be very red (u-g $>$ 0.8), strong IR
emitters (\citealt{LPM93}; for a detailed study of such a case see
\citealt{Veron06}) and are not suitable for the FeII template
adopted for this study. We identified three RL sources whose spectra
show extreme FeII emission (R$_{FeII}$ much larger than 2.0): SDSS
J094927.67+314110.0, SDSS J144733.05+345506.7, SDSS
J152350.42+391405.2. Our attempt to fit the IZw1-based
(\citealt{BG92}) template for such objects was unsuccessful, thus
they are not shown in Figure 4. Such objects require special
attention and is beyond the purpose of the present study to focus on
their nature. The inclusion of such pathological and relatively rare
cases will not affect the conclusions of the present study. We were
able to make reliable measures in the H$\beta$ region for N=477
objects. Our RL/RI sample is complete to 17.5 apparent magnitude in
g- OR i-band. We are confident that the RL+RI sample is at least
75\% radio-complete because all have a FIRST S(1.4GHz)$\geq$ 1.5 mJy
(see Figure 1 of \citealt{Jiang07}). Our RL sample includes n=85 RL
quasars (n=46 FRII) and n=59 RI QSOs. The remaining n=333 represent
our RQ sample which, while incomplete, is large enough to be
representative for the RQ parent population. Our final sample
includes sources spanning the extinction corrected i-band range
-27.5$\leq$ M$_{i}$$\leq$-17.1. We cross-checked our entire sample
selected based on psf g $<$ 17.5 (806 ``QSO'' + 16 ``Galaxy''- see
Tables 1 and 2) with the ``vetted'' QSO catalog \citet{Schneider07}.
All sources in our sample with M$_{i}$ brighter than -22.0 are
present in the that catalog. All sources in that catalog satisfying
our selection criteria are found in our sample. No additional QSO
satisfying our selection criteria was found there.

\section{Analysis of Optical Spectra from SDSS}

In order to obtain the optical parameters of the 4DE1 we followed
the analysis procedure described in \citet{Marziani03a}\footnote{We
used Image Reduction and Analysis Facility (IRAF), distributed by
the National Optical Astronomy Observatories, which are operated by
the Association of Universities for Research in Astronomy, Inc.,
under cooperative agreement with the National Science Foundation -
http://iraf.noao.edu/}. An underlying power-law
$f_{\lambda}\sim\lambda^{\alpha}$ continuum was defined using
regions minimally contaminated by FeII lines, specifically at
4195-4215 {\AA} and 5700-5800 {\AA}; it was decided prior to fitting
the FeII template. We used the IZw1-based template of \citet{BG92}
for FeII decontamination. This represents an important advantage
over our own Atlas sample \citep{Marziani03a} where the typical
wavelength coverage of the spectra was $\sim$1000\AA, rendering
continuum estimation very uncertain. Our chief goals from FeII
template fitting are: 1) to obtain W(FeII 4750{\AA}) blend and 2) to
clean up the H$\beta$ region.

In about 20\% of sources we attempted to remove the host galaxy
contamination\footnote{When prominent absorptions lines like Mg
$\lambda$5177{\AA} and Na $\lambda$5896{\AA} are present} before
deriving 4DE1 parameters using the library of theoretical galaxy
templates from
GALAXEV\footnote{http://www.cida.ve/$\sim$bruzual/bc2003} of
\citet{BC03}. A much more sophisticated approach is proposed by
\citet{Zhou06}. All our sources with noticeable host galaxy
contamination lie within z=0.2. There seems to be an apparent
contradiction with the results of \citet{Vandenberk06}. They use an
eigenspectrum decomposition technique and report a non-zero host
galaxy contamination all the way to z=0.7. Our sole purpose was to
clean up the spectral region of interest of the most prominent
absorption lines of a host galaxy. We make no attempt to estimate
the relative proportion of AGN and host light in the spectrum unless
clear absorption lines are observable; the spectra have good S/N to
reveal potential host contaminations. Evidently, most objects we
deal with tend to have bluer continua (see Figure 1 and the previous
section that explains how we built-up our sample) and thus are less
affected by host galaxy. It is possible that many of the RQ sources
not considered for our spectroscopic reduction have a significant
host galaxy contribution at redshifts higher than 0.18. We are also
aware that we severely under-represent the population of low z
objects that form the ``bump'' in Figure 1. As we try to explain in
this section, the FWHM H$\beta$ seems to be minimally responsive to
this uncertainty in the relative amount of flux from a host galaxy.
See Figure 3 for an example of the host galaxy ``removal''. In most
cases this process turned out to be very effective in revealing the
QSO emission line spectra. We tested the effect of host galaxy
subtraction on our spectroscopic measures and found that $\sim$ 60\%
of these sources (that required this step) were seriously
contaminated and the rest moderately/weakly contaminated. For the
latter, we attempted an estimate of the 4DE1 optical plane both
before and after the correction. We found that FWHM H$\beta$ changes
randomly within $\pm$5\%, but R$_{FeII}$ was much more sensitive to
this procedure, with a 25-35\% systematic change toward lower
R$_{FeII}$ values. Our set of contaminated spectra span a wide range
of 4DE1 parameters values.

\begin{figure}
   \plotone{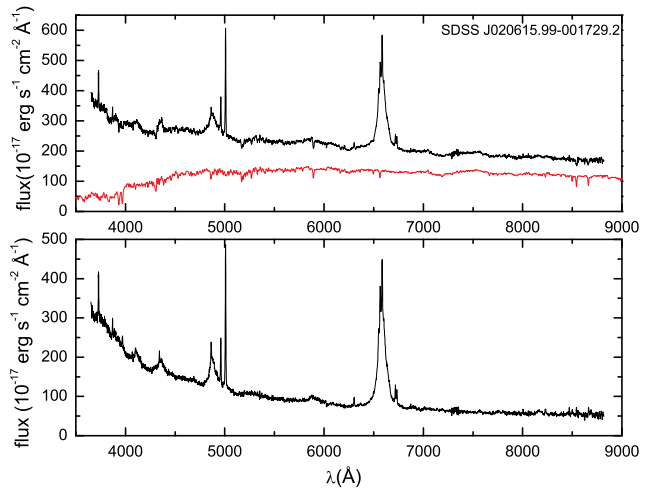}
   \caption{An example
of a QSO contaminated spectrum. The top panel shows the initial SDSS
spectrum (black) and a template of a host galaxy (red); the bottom
panel shows the QSO spectrum after subtracting the spectrum of the
host galaxy.}
\end{figure}

When clear inflections were observed between the broad and narrow
components of H$\beta$, we did not constrain the width of the latter
component to be the same as FWHM [OIII]$\lambda$$\lambda$ 4960,
5008\AA. No attempt was made to decompose the broad Balmer line into
components. A spline function was used to fit its global profile.

We compare the values obtained in the present study for FWHM
H$_{\beta}$  and R$_{FeII}$ with those measured in
\citet{Marziani03a} for the n=38 sources in common with that Atlas.
We calculated $\Delta$FWHM H$\beta$ and $\Delta$R$_{FeII}$ for every
object (of these 38) considering in each case the values obtained in
the current study and those in the Atlas (2003). The mean and median
differences are a reasonable indicator of the robustness of the 4DE1
parameter space in its optical dimensions. We find a scatter of
$\sim$ 10-20\% for FWHM H$\beta$ (no systematic effect) and for
R$_{FeII}$ we report a systematic effect of 30-35\% toward larger
values in our present study. This latter effect may be due to a more
reliable choice of the continuum and/or a higher quality of spectral
signal. For Figures 4 and 5 we conservatively adopt the
uncertainties estimated in the Atlas (2003), even though the quality
of the spectra is clearly better in our present sample. We are aware
that the R$_{FeII}$ gap (c.f. Figure 4) between 0 and 0.1 is not
physical and most likely reflects the difficulty of measuring very
low values of W(FeII $\lambda$4570\AA). Such an effect could be due
at least in part to our simple definition the optical continuum. We
would like to explore more on this issue in a future project.

\section{Locus of RL and RI Quasars in the 4DE1 Optical Plane}

The optical plane of 4DE1 provides a powerful diagnostic tool for
testing whether the RL and non-RL sources are spectroscopically
different. We earlier proposed the existence of two QSO populations
A and B separated at FWHM H$\beta$$\simeq$ 4000 km s$^{-1}$ (see the
Introduction and section 3.2 of \citealt{Sulentic07}; see also
section 3 of \citealt{Sulentic00a}).

The previously reported restricted domain occupation of RL sources
in 4DE1 space means that we can now ask if: 1) the SDSS sample
confirms the earlier restricted RL domain occupation and 2) if  RI
sources show domain occupation more similar to RL or RQ Type 1 AGN.
Figure 4 shows the distribution of RL and non-RL sources in the
optical plane of 4DE1 (non-RL include n=333 RQ + n=59 RI). Figure 4
clearly confirms a restricted domain space occupation for RL sources
with $\sim$ 78\% falling above FWHM H$\beta$=4000 km s$^{-1}$ (our
so-called population B domain: \citealt{Sulentic00a}); 91\% of FRIIs
and 62\% of CD RL quasars are in this Population B. The horizontal
line in Figures 4 and 5 marks this nominal populations boundary. RQ
QSOs show a much wider domain space occupation with more than half
($\sim$ 62\% of our sample) lying below FWHM H$\beta$=4000 km
s$^{-1}$. If 4DE1 parameters measure aspects of Broad Line Region
(BLR) kinematics and geometry then this domain occupation difference
is consistent with a physical difference between RL and the majority
of RQ sources, which supports a RQ-RL bimodality. At the very least
past work reported that RL sources show systematically higher black
hole masses and systematically lower Eddington ratios than the RQ
majority \citep[e.g.][]{Marziani01, Boroson02, Marziani03b,
Sulentic06}.

We performed a 2D Kolmogorov-Smirnov (K-S) test in order to evaluate
the significance of: 1) the RQ-RL difference and 2) the RL FRIIs -
RL CDs difference in domain occupation. Following the routine
available in Numerical Recipes\footnote{www.nr.com} the K-S
procedure (\citealt{Peacock83,FF87}) divides the optical plane into
quadrants that maximize the two population difference. For test 1)
the RQ sample is assumed to represent the parent Type 1 AGN
population and RL the test sample. For test 2) the RL FRII sample is
assumed to represent the parent population and the RL CDs the test
sample. The results of the two tests are summarized in Table 3. One
can notice that the RL/non-RL separation at FWHM H$\beta$ = 3875 km
s$^{-1}$ is in reasonable agreement with our previously adopted
Population A/B boundary at 4000 km s$^{-1}$ (Figure 4). As reported
in Table 3, the probability that RL and RQ occupy the same
spectroscopic domain is very low. Similarly, the second test between
FRIIs and RL CDs shows that the two samples are very distinct in
terms of spectroscopic properties. In the former case the result is
equivalent to saying that it is extremely unlikely that RL and
non-RL are drawn from the same parent population. In the later test,
the probability listed in Table 3 could be interpreted in two ways:
1) the orientation is responsible for the distinct space occupation
for FRIIs and RL CDs or 2) the CD RLs and the FRIIs are drawn from
distinct populations, in which case the RL CDs (all or most of them)
could be interpreted either as beamed RQs or as pre-/postcursors of
an FRII episode.

\begin{table*}
\caption{Two-dimensional Kolmogorov-Smirnov tests for RL / non-RL
and for RL FRII / RL CD.} \tabcolsep=3pt
\begin{tabular}{lcc}
\hline

Samples & Coordinates of quadrants & Probability  \\
n1-parent and n2-test & FWHM H$\beta$ (km s$^{-1}$) ; R$_{FeII}$ &
of null hypothesis \\

\hline\hline

392-non-RL and 85-RL & 3875 ; 0.49 & P$\sim$6.2$\times$10$^{-8}$ \\

\hline

46-RL FRII and 39-RL CD & 6100 ; 0.18 & P$\sim$9.8$\times$10$^{-4}$ \\

\hline

\end{tabular}

\end{table*}

There is also a clear bimodality of RL/RQ in terms of R$_{FeII}$.
(robustly confirmed by the 2D K-S test we performed). In the context
of 4DE1 we focus on FWHM H$\beta$ for several reasons: 1) it is a
direct measure of the Broad Line Region (BLR) kinematics, 2) it can
be measured more accurately than R$_{FeII}$ (see \S~3) and 3) there
are significant Population A/B differences reported over the last
seven years that are defined in terms of FWHM H$\beta$ alone (see
Table 5 in \citealt{Sulentic07}).

\begin{figure}
\plotone{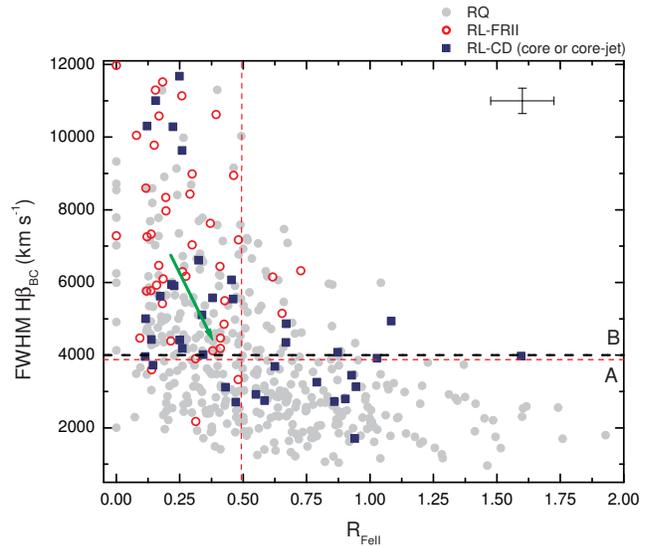} \caption{RL quasars in the optical plane of the
4DE1 parameter space. The green arrow indicates the displacement
between the median FWHM H$_{\beta}$ and the median R$_{FeII}$ for
the FRII and the CD sources. The solid light gray symbols are non-RL
objects. In the upper right corner are indicated the typical
2$\sigma$ errors, estimated in \citet{Marziani03a}. The red dotted
lines show the boundaries for the quadrants defined by the 2D K-S
test we performed as we explained in the text. The black dotted line
indicates the Population A/B boundary. The vertical axis is
truncated at 12000 km s$^{-1}$ for clarity, thus we miss showing
five other RL (two CD and three FRIIs between 12000-24000 km
s$^{-1}$).}
\end{figure}

Figure 4 then shows a significant displacement between the non-RL
and RL distributions with most RL lying above FWHM H$\beta$=4000 km
s$^{-1}$. We also observe a separation between the mean/median
position of FRII RL sources and core/core-jet (CD) RL sources (see
Table 4), confirming a result from \citet{Sulentic03}. This is the
first step in estimating the role of source orientation in 4DE1. The
vector shown in Figure 4 indicates the change in median 4DE1 optical
parameters between FRII and CD RL sources. Orientation-unification
scenarios see the latter sources as having radio jets aligned to
within a few degrees of our line of sight. The vector suggests that
source orientation strongly influences FWHM H$\beta$ measures and,
to a lesser degree, R$_{FeII}$. Given the likelihood that large
disk-jet misalignments can occur and that radio structure in many RL
is highly nonlinear it is surprising how large is the FRII-CD median
parameter separation. The 10-20\% of CD and FRII RL with,
respectively, very large and very small FWHM values could be
interpreted as sources where the radio structure and the region that
emits the broad lines (i.e. accretion disk) appear to be misaligned,
if one invokes the unified picture for AGNs relative to Figure 4.
The few RL CDs with very large FWHM H$\beta$ may, apparently
disconnected from the bulk of RL CDs, may be the best candidates for
a pre- or post- FRII phase. Sources with extremely broad Balmer
profiles (sometimes double peaked) are so rare (we find a handful of
such sources with FWHM H$\beta$ in the range 12000 - 30000 km
s$^{-1}$) as to defy interpretation as the simple tail of a normal
QSO FWHM distribution.

Figure 4 included RI along with the RQ QSOs as non-RL QSOs. The
tacit assumption was that RI and RQ are the same. Figure 5 presents
a test of that assumption that is equivalent to that performed for
RL. Does the previously defined RI sample show a distribution in the
optical plane of 4DE1 more similar to RL or RQ sources? The n=59 RI
sources (stars in Figure 5) show no distinguishable difference in
occupation from the n=333 RQ sample. Only 42\% of the RI population
is found in the Population B domain, comparable to the $\sim$ 37\%
for the RQ sources. There is therefore no spectroscopic evidence
that the RI sources form a special class (see Table 4).

\begin{table*}
\caption{Mean and Median optical spectroscopic measures for the
samples used in Figures 4 and 5.} \tabcolsep=3pt
\begin{tabular}{lcccc}
\hline

            & \multicolumn{2}{c}{mean$\pm$standard deviation} &
\multicolumn{2}{c}{median} \\

\cline{2-5} \\

           & FWHM H$\beta$ & R$_{FeII}$ & FWHM H$\beta$ & R$_{FeII}$ \\

           & (km s$^{-1}$)& & (km s$^{-1}$) & \\

\hline\hline \multicolumn{5}{l}{Relative to Figure 4} \\
\hline

RL (n=85) & 6809$\pm$3976 & 0.36$\pm$0.29 & 5775 & 0.27\\
non-RL (n=392 RQ+RI) & 4016$\pm$2569 & 0.57$\pm$0.38 & 3375 & 0.50\\

\hline

RL FRII (n=46) & 7673$\pm$3733 & 0.26$\pm$0.17 & 6750 & 0.20\\
RL CD (n=39) & 5789$\pm$4059 & 0.48$\pm$0.36 & 4418 & 0.38 \\

\hline \hline \multicolumn{5}{l}{Relative to Figure 5} \\
\hline

RQ (n=333) & 3852$\pm$2114 & 0.56$\pm$0.35 & 3227 & 0.50\\
RI (n=59) & 4941$\pm$4230 & 0.62$\pm$0.52 & 3659 & 0.50\\

\hline

\end{tabular}

\end{table*}

\begin{figure}
\plotone{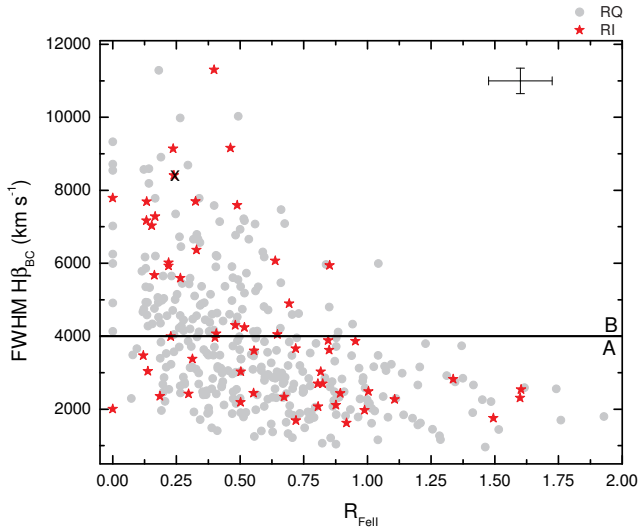}\caption{RI QSO in the optical plane of the 4DE1
parameter space. The light gray symbols show the RQ objects and the
solid red stars are the n=59 RI sources. SDSS J232721.97+152437.3 is
indicated with an ``X'' - see ``Discussion'' and also Table 3. The
vertical axis is truncated at 12000 km s$^{-1}$ for clarity, thus we
miss showing two other RI.}
\end{figure}

\section{Can We Reveal a RL/RQ Dichotomy Using L-dependent Diagrams?}

The previous section compared RQ, RL and RI sources in a
Luminosity-independent context. We now address the problem of a
RL/RQ dichotomy from a Luminosity-dependent perspective. Figure 6
plots source bolometric versus radio luminosity (L$_{bol}$ vs.
L$_{1.4GHz}$). The radio luminosity is K-corrected
(\citealt[]{Hogg99}) assuming that f$_{\nu}$ $\sim$ $\nu$$^{\alpha}$
and $\alpha$ = - 0.5 in the radio regime. The bolometric luminosity
was estimated from L$_{bol}$$\simeq$10$\lambda$L$_{\lambda}$, where
$\lambda$ $\equiv$ 5400\AA\ \citep[see a concise discussion on the
bolometric correction in section 2.8 of][and references
therein]{Marziani06}. The 5400\AA\ specific luminosity is estimated
using the continuum flux in the rest-frame spectrum of the QSO. We
used \textit{dopcor} task in IRAF with the the appropriate
cosmological flux corrections applied when deredshifting the
spectra. Our sample covers about four decades in bolometric
luminosity with logL$_{bol}$ $\sim$ 43.0-47.0 erg s$^{-1}$.

\begin{figure}
\plotone{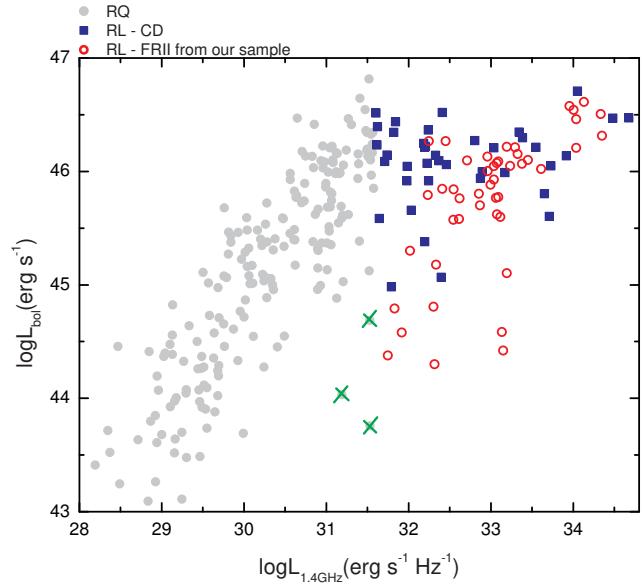}\caption{The distribution of our relatively
bright sources in a plane defined by the bolometric and radio
luminosity. The three objects marked with an X are commented in
Table 3 and related discussion in \S~\S~8.1. The radio-undetected
QSOs are not shown here.}
\end{figure}

There are a few important comments about Figure 6: a) we find no RL
quasar fainter than logL$_{bol}$ = 44.3 erg s$^{-1}$, b) we find no
CD RL below logL$_{bol}$ $\simeq$ 45.0 erg s$^{-1}$, c) the RQ
sample shows a power-law correlation (L$_{bol}$ $\propto$
L$_{1.4GHz}$$^{0.89}$) (see also e.g. \citealt{Kukula98}) or
alternatively L$_{1.4GHz}$ $\propto$ L$_{bol}$$^{0.85}$ (the same as
reported in \citealt{White07}, considering that L$_{bol}$ $\propto$
L$_{opt}$) but d) the RL population however shows a different
behavior for CDs and FRIIs. The majority of CDs concentrate at high
values of L$_{bol}$, while the distribution of FRII quasars shows a
rough correlation parallel to the RQ one. Most of the CDs,
especially below logL$_{1.4GHz}$ $\simeq$ 32.5, make no sense in an
orientation unification scenario because there is no corresponding
weaker FRII population to the left of them from which they could be
boosted. As we suggested in \S~4 based on the 2D K-S test for
FRIIs/RL CDs these sources could be seen as boosted RQs.

4DE1 parameters show no obvious dependence on source luminosity.
They also show no dependence on radio luminosity apart from a
restricted domain space occupation observed only for RL sources. If
RQ and RL sources belong to the same family then we could reasonably
expect them to follow the same correlation between bolometric and
radio luminosity. Figure 6 suggests that RQ and RL sources show
separate correlations. We see a clear {\it dichotomy} between the
populations at our previously determined boundary
(logL$_{1.4GHz}$=31.6 erg s$^{-1}$ Hz$^{-1}$) as lower limit for RL
activity. This dichotomy appears to independently confirm our
previous suggestion, based on 4DE1 occupation, that RQ and RL
sources are fundamentally different.

Figure 6 is based upon our bright SDSS sample. Does it include all
FRII quasars within z=0.7? What is the effect on Figure 6 of
including fainter sources in the same redshift range? Consideration
of these questions can help us understand why such conflicting
results about a RQ-RL dichotomy/bimodality have been obtained in
past studies. We have reason to fear that Figure 6 does not tell the
full story about dichotomy because almost all sources in our sample
with logL$_{bol}$$<$45.5 erg s$^{-1}$ show z$\leq$0.15 while all
sources above that value show z$>$0.15 (see Figure 8). We have
essentially sampled the bright end of the low redshift Optical
Luminosity Function (OLF). On this bright end RL sources are
relatively abundant and RQ numbers small enough to allow a dichotomy
to be seen. But we have severely undersampled the faint end of the
OLF \citep[e.g.][]{Boyle00, Croom04, Richards05, Richards06}. In
that luminosity range the RQ population is so large that the radio
bright tail of the RQ distribution might overlap the RL distribution
effectively quenching any dichotomy.

The above suggestion can be tested and illustrated by adding fainter
subsamples of QSOs (see Table 2) to Figure 6, leading to Figure 7,
which shows the following : i) the n=67 sample of FRII from
\citet{deVries06}, fainter than our sample of FRII (see \S~2); ii)
all radio-detected objects that are labeled ``Galaxy'' by the SDSS
spectro-pipeline but show Type 1 spectra (no spectroscopic reduction
was performed on these, as we explained earlier); iii) core radio
sources (no FRIIs) with psf g in the range 19.0-19.5 and iv) core
radio sources (no FRIIs) with psf i in the range 19.0-19.5. In order
to understand some of the the subtle effects that come into play at
this point Figure 7 should be approached in conjunction with Figures
8 and 9. We point out that for Figures 7 and 8 we estimated the
bolometric luminosity following the empirical results of
\citet{Hopkins07} - first estimating the B-band luminosity for our
sources obtained from psf u and psf g magnitudes (corrected for
extinction, using the SDSS coefficients). The B-band luminosity
(K-corrected, assuming that f$_{\nu}$$\sim$$\nu$$^{\alpha}$, where
$\alpha$=-0.5 in the optical domain) was obtained from psf u and psf
g magnitudes using the transformation formula proposed by
\citet{Jester05a}. We make no attempt to reduce the spectra of these
faint objects and therefore no 5400\AA\ specific luminosity (in the
underlying continuum) can be estimated and used to estimate
L$_{bol}$. This is why we employ the the empirical recipe from
\citet{Hopkins07}. It is worth mentioning that we examined every
optical spectrum and radio image in FIRST/NVSS to confirm Type 1
optical and FRII/CD radio status. No attempt was made to estimate a
host galaxy contribution for these fainter sources. The main results
of this study (e.g. Figures 6, 7 and 8) are not sensitive to the
choice of the bolometric correction. We also remind the reader that
the distribution in Figure 9 reflects the heterogeneous construction
of the radio sample we investigated, as explained in the previous
sections.

\begin{figure}
\plotone{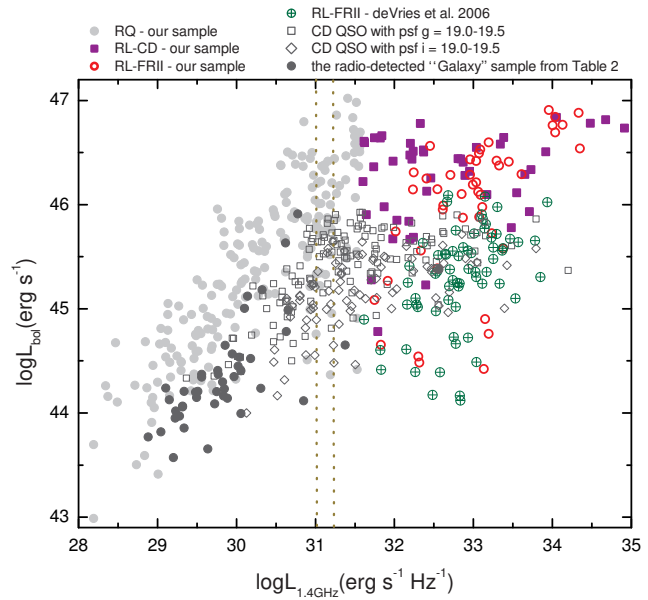}\caption{The distribution of our relatively
bright sources plus several subsamples of QSOs selected as explained
in the text. The radio-undetected QSOs are not shown here. The
vertical dotted lines at 31.0 and 31.2 represent the minimum
detectable radio luminosity that corresponds to a detection limit of
0.7 - 1.0 mJy at z=0.7, respectively.}
\end{figure}

\begin{figure}
\plotone{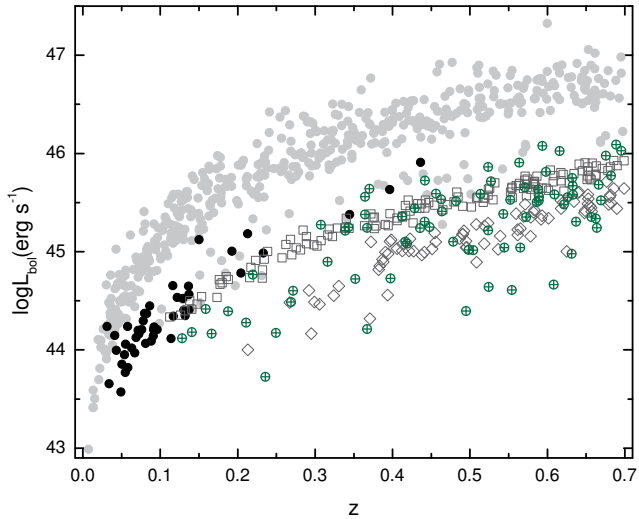}\caption{The luminosity distribution with
redshift for the various subsamples of QSOs employed in Figure 7;
the symbols are similar to those in Figure 7, with the exception
that this time all QSOs in our sample (RQ, RL CDs and RL FRIIs) are
displayed with the same solid light grey symbols. The
crossed-circles denote only the FRII subsample from
\citet{deVries06}, just as in Figure 7.}
\end{figure}

\begin{figure}
\plotone{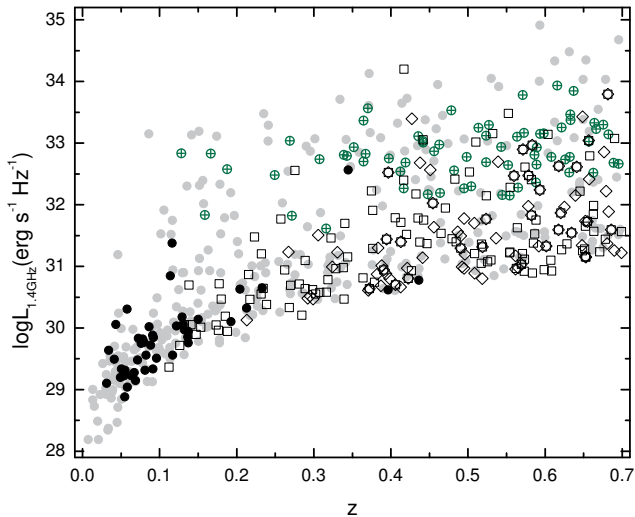}\caption{The radio luminosity distribution with
redshift for the various subsamples of QSOs employed in Figure 7 and
8; the symbols are identical to those in Figure 8.}
\end{figure}

In Figure 7 it is now obvious that the optically faint radio core
sources (the empty squares and diamonds) tend to fill in the region
logL$_{bol}$ $\sim$ 44.0 - 46.0 and logL$_{1.4GHz}$ $\sim$ 30.5 -
32.0. The fact that such objects are less luminous than logL$_{bol}$
= 46.0 is not a surprise. They are selected as apparently fainter
objects in the same z range as the bright sample. It is important
noting that these fainter samples (iii and iv above) are not
represented below z=0.1 and z=0.2, respectively. This is the main
cause of their tendency to scatter mostly in the region that was
previously (in Figure 6) scarcely populated along the
radio-luminosity axis. The main conclusion we get is that the
picture of the dichotomy becomes ``blurry'' at this point.

With the addition of the fainter FRIIs from \citet{deVries06} we
observe the parallel RL FRII sequence more clearly in Figure 7. The
trend we see for FRII quasars is qualitatively similar to that
reported by other studies for steep-spectrum quasars using lower
radio frequencies (e.g. \citealt{Serjeant98, Willott98}) or higher
radio frequencies \citep{Xu99}. The two correlations likely indicate
a different fraction of power output channeled into the jets
\citep[e.g.][]{RS91, Miller93}. The core of RL population covers
$\sim$2.5dex in the L$_{bol}$ space and $\sim$2dex in the
L$_{1.4GHz}$ space, while the RQ sample distributes over $\sim$4dex
in either measure. The small number of extreme sources above
logL$_{bol}$$\sim$46.5 lie near our high redshift limit and suggest
that we are seeing the first hints of luminosity evolution in our
sample.

\section{Estimating the Radio Loud Fraction (RLF) of Type 1 AGN}

We are now in a position to estimate the fraction Radio Loud quasars
relative to the total population of QSOs that satisfies our redshift
and luminosity criteria. Figure 1 shows that an i-band selection
adds a large number of relatively low redshift sources. Bluer
(g-band) selection provides a more uniform sampling of sources in
our redshift range. If we consider only sources with psf g $<$ 17.5
the RLF $\simeq$ 9.2\% (of n $\sim$ 822 sources, i.e. 806 QSO + 16
``Galaxy'' - see Tables 1 and 2). Considering sources selected with
psf i $<$ 17.5 yields a RLF $\simeq$ 5.2\% (relative to n $\sim$
1737 objects, i.e. 1656 ``QSO'' + 81 ``Galaxy'' - see Tables 1 and
2). If we combine the above conditions with a logical ``OR''
operator we get RLF $\simeq$ 5.4\%, relative to n $\sim$ 1770
objects.

Our results are quite different compared to the RLF estimate for the
PG sample of QSOs; RLF $\sim$ 17\% (applying our definition of
radio-loudness to the sample of 87 sources in \citealt{BG92} or
directly to the full list of 96 Palomar-Green UV-excess selected
QSOs/Seyferts from \citealt{GSL86} with z $<$ 0.5). \citet{BG92}
include 50 Population A sources (two of them RL) and 37 Population B
sources (twelve of them RL). Obviously, one must also bear in mind
that the absolute magnitude cut at M$_{B}$=-23 in the PG sample
\citep{BG92} has its play into the rather large RLF it includes,
considering also our results presented in Figure 6; we find no RL in
the dimmest decade of bolometric luminosity we sampled. This is also
consistent with the reported dependence of RLF on luminosity (see
the ``Introduction'' for references related to this issue).

\citet{Jester05a} offered a detailed discussion on possible
radio-related incompleteness in the Palomar-Green Bright Quasar
Survey (\citealt{SG83,GSL86}) compared to an i-band limited sample
from the SDSS. They suggest that the rather large RLF in the BQS
appears connected to the fact that BQS objects, being selected in a
B-band flux-limited survey, have rather blue continua, objects with
bluer continua apparently tend to have stronger [OIII] lines, and
objects with larger [OIII] lines are more likely to be radio-loud.
In the present study we also find a RLF fraction approximately
double when selecting the sample based on a bluer filter (g-band)
than when selecting based on a redder one (i-band). On the other
hand \citet{Jester05a} find no systematic radio-related biases by
comparing the BQS sample against a BQS-like sample selected from
SDSS database. More recently, two other SDSS/FIRST - based studies
\citep{deVries06, Lu07} found that FRII quasars are rather bluer
than the radio-compact sources, which apparently could be connected
to the RL excess in BQS survey. Indirectly confirming their
conclusion (i.e. corroborated with our results that the large
majority of FRIIs are members of Population B) is the study of
\citet{Richards03}, which report a systematic narrowing of the
H$\beta$ line with increasing redness.

However, the cause of the RL excess in PG survey is not completely
clear at this time, considering also the apparently contradictory
conclusions about the optical colors of radio quasars, i.e. RL
quasars have been found to be in general redder than RQ QSOs (e.g.
\citealt{Brotherton01, Ivezic02, White07, Labita07}).

\section{4DE1 optical plane and the Radio-Loudness Probability}

4DE1 is a diagnostic tool that could set empirical constraints for
the theoretical models of AGN physics. Our results show that RL
quasars prefer a restricted zone of occupation in the optical plane
of 4DE1 (Figure 4). Our g-band selected sample of n=333 QSOs (n=34
RL) is the most complete that we have available for this analysis.
The fractions of Population A and B sources are $\sim$60\% and
$\sim$40\% respectively of which 4-5\% and 17\% are RL. The
situation is even more extreme if we correct sources for
line-of-sight orientation (see Figure 4). In an orientation
unification scenario many of the Population A RLs (mostly CDs) could
simply be face-on oriented Population B RLs. FRII sources can be
said to define the locus of the RL population in the 4DE1 optical
plane. If one extrapolates the relative proportion of Population A/B
sources to the whole sample of n=1770 (see the previous section),
the radio-loudness probability would be 10\% and 2\% for Population
B and A respectively. A quasar is approximately 4$\times$ more
likely to be RL if it shows a Population B optical spectrum. The
scarcity of RL quasars in the population A domain is not in dispute.
Recent studies \citep{Komossa06} searching for RL Narrow-Line
Seyfert 1 (NLSy1) sources (extreme population A QSOs with FWHM
H$_{\beta}$ $\leq$ 2000 km s$^{-1}$) find that most so-called RL
NLSy1 are in fact RI QSOs. A very small number of {\it bona fide} RL
NLSy1 is known at this time \citep[e.g.][]{Zhou02, Zhou03, Zhou06,
Komossa06}.

\section{Discussion}

One of the most fundamental differences among the broad line QSOs
involves the existence of RQ and RL populations. Do all sources pass
through a RL phase? Do RL quasars represent in some way a physically
distinct class? The latter question motivates the high level of
interest in the possibility of a RQ/RL dichotomy. We have shown that
plots in terms of radio and bolometric luminosity (Figures 6-8) are
not the best way to answer it. Samples restricted to high L$_{bol}$
sources show hints of a dichotomy, but more complete samples do not.
This is due to the numerical imbalance in the two populations. A
complete sample of QSOs contains so many RQ sources that the radio
brightest of that population will bridge the gap between RQ and RL.

4DE1 provides a better way to address the problem. A way that is
independent of the radio and optical luminosity of sources. 4DE1
suggests that the answer to the latter question may be ``yes''. If
4DE1 parameters measure fundamental aspects of BLR kinematics and
geometry then we have evidence that the RL quasars (mostly
Population B) may be significantly different (\S~4) at a fundamental
level from the majority of RQ QSOs which occupy the Population A
domain. Taken at face value, the 4DE1 optical plane suggests
necessary (yet not sufficient) empirical constraints for developing
RL activity. It is important to remember that the RQ-RL separation
in 4DE1 is not complete. About 60\% of RQ QSOs (Population A) show
properties almost never seen in RL sources while about 40\% of RQ
sources are spectroscopically indistinguishable from RLs (Population
B). In this case Population B RQ sources apparently have the
necessary BLR properties for radio loudness but not sufficient to be
RL.

One possibility is that population B RQ (and especially RI) sources
might be the pre- or post-cursors of the RL phase. However, another
scenario is that the population B RQ sources occupying the RL domain
simply reflect the overlap of two unrelated AGN sequences. The idea
of two distinct populations suggests that something else is a
necessary ingredient in AGN physics that manifests/triggers radio
loudness; it has been suggested the BH spin \citep[e.g.][]{WC95,
Moderski98, Meier01, Volonteri07}, the host galaxy morphology
\citep[e.g.][]{Capetti06, Sikora07} and/or its link with the nucleus
(e.g. \citealt{Hamilton08}), the environment
\citep[e.g][]{Kauffmann07}. Some more clues could come from an
analogy with X-ray binaries (e.g. \citealt{Maccarone03, Jester05b,
Kording06a,Kording06b}). Those studies suggest that RL quasars are
in a distinct accretion mode compared to RQ QSOs. Moreover, for a
better understanding of radio-loudness one should also consider the
ratio of optical:X-ray emission (i.e. disk:corona relative
emission)(\citealt{Kording06b}). Future studies can certainly
explore these valuable arguments incorporating the empirical data
presented here.

The overlap of RL and RQ sources in the Population B domain suggests
that population A-B distinction may be more physical than RQ-RL
comparisons (see also \citealt{Boroson02}). The two populations (A
and B) are nominally separated at FWHM H$\beta$=4000 km s$^{-1}$ and
there are many other forms of evidence that support a boundary near
this value \citep{Sulentic07}. We suggested that this might
correspond to a critical Eddington ratio (log L/L$_{Edd}$$\sim$0.15)
where the BLR properties change rather suddenly
\citep[e.g.][]{Sulentic00b, Marziani01, Marziani03b, Marziani06,
Sulentic07}. In a recent study \citep[][with reference to
\citealt{Bonning07}]{Kelly08} argue that L$_{bol}$/L$_{Edd}$
$\approx$ 0.3 could indicate some critical change in the accretion
disk structure. At this time we can only add that all objects in our
sample that have L$_{bol}$/L$_{Edd}$ larger than this value are
exclusively part of the Population A while all others showing
L$_{bol}$/L$_{Edd}$ less than 0.3 are a mixture of Population A and
B.

Comparing median FWHM H$\beta$ values for FRII and CD RL sources
gives 6750 and 4400 km s$^{-1}$ respectively. This difference is
interpreted as a manifestation of source orientation. CDs viewed as
near disk face-on (alternatively jet-aligned) sources show FWHM
measures that are not dominated by Keplerian motions in contrast to
FRII quasars. Detection of this FRII-CD difference in median FWHM
\citep[see also][]{Rokaki03, Sulentic03, deVries06} supports BLR
models involving a flattened, disk-like geometry (for a more
detailed discussion on BLR structure and dynamics see section 3.1 in
\citealt{Collin06}). The results on the relative distribution of RQ,
as well as FRII and CD RLs, in Figure 4 indicate that face-on RL
sources contain a significant extra ($\sim$2-3000 km s$^{-1}$)
component of line-of sight motion that is not present in face-on RQ
AGN. That may or may not be associated with the radio jets.

\subsection{Radio Intermediates}

Armed with evidence that BLR structure in RQ and RL sources may be
fundamentally different we return to the sources with log
L$_{1.4GHz}$$\sim$31.0-31.6 erg s$^{-1}$ Hz$^{-1}$. They are one, or
some combination, of the following: a) radio-weakest RLs, b)
radio-strongest RQs or c) a special class (RI) of sources. We
disfavor interpretation a) because we find no bona-fide FRII radio
morphologies among them. We favor option b) because they show CD
emission (like RQs), but cannot be boosted FRIIs. Claims of
relativistic jet detection \citep[e.g.][]{Blundell98} in some of
these quasars has led to the suggestion that they might be boosted
RQ (option b) sources \citep[e.g.][]{Miller93, Falcke96a, Wang06},
which would even more effectively bridge the gap between the
majority of RQ sources and the RL quasars. Since the RI show no
obvious difference from weaker and unambiguously RQ sources we
disfavor option c) and again favor option b).

It has also been proposed that the radio emission in RQs (or
alternatively, non-RLs) is mostly related to star formation
processes (circumnuclear starbursts, supernovae)
\citep[e.g.][]{SA91, Terlevich92, Miller93, CP95}. This idea appears
to be naturally related to the fact that RQ sources follow the
radio-FIR correlation, as mentioned earlier. On the other hand, the
discovery of flat radio spectra, elongated radio cores in non-RL
quasars and/or high brightness temperatures
\citep[e.g.][]{Falcke96a, Blundell98, Wang06, Leipski06} favors the
hypothesis that the mechanism of radio emission in non-RLs is
similar to that of RLs. However, the claim of a relativistic jet in
``RQ'' PG1407+265 \citep{Blundell03} involves an unambiguously RL
quasar by our definition (log L$_{1.4GHz}$=32.5 erg s$^{-1}$
Hz$^{-1}$). Another deep search \citep{Ulvestad05} failed to confirm
some of the other detections in \cite{Blundell98} indicating that
the frequency of occurrence of weak jets in RQ quasars is still
uncertain. Some of the claims involve AGN that do not show Type 1
spectra \citep[e.g.][]{Falcke00, Nagar00, Nagar01, GB02}, which are
not considered here.

All RI sources show core (or core-jet) morphology leading us to
conclude that their radio emission may be fundamentally different
from the classical RL sources (option c?). They could be frustrated
jets that face too much resistance from the ambient medium, thus
failing to manifest as large scale FRII structures. They cannot be
boosted classical RLs unless we have missed a \textit{significant}
unboosted FRII population which would presumably lie in the zone of
our RI sample. Our comparison of NVSS and FIRST fluxes for the RI
sample does not allow us to rule out the existence of hidden FRIIs.
A total of n=37 RI were detected in both radio surveys and we find
about 18 QSOs with an NVSS flux density excess in the 10-50\% range.
Some of these involves unrelated point sources within the NVSS beam.
A complementary approach is to look for unusual radio structures in
the FIRST and NVSS maps.

Table 3 summarizes the properties of quasars showing unusual radio
structure. Three of these sources likely involve weak FRII structure
while the rest show no hint of it. These three sources are marked
with an X in Figure 6. Their location along the trend described by
the bona-fide FRIIs increases the likelihood that they may be very
weak FRIIs.

\begin{table*}
\caption{Radio Quiet and Intermediate sources in our sample that
show peculiar radio morphology.} \tabcolsep=3pt
\begin{tabular}{lc}
\hline

Name & Comments \\
\hline\hline

SDSS J110717.77+080438.2 & galaxy nearby; core-jet morphology\\
SDSS J114047.9+462204.8 & misaligned nearby radio structure with no\\
 & SDSS-detected counterpart \\
SDSS J120014.08-004638.7 & possible core+lobes structure \\
SDSS J171322.58+325627.9 & possible lobes \\
SDSS J230443.47-084108.6 & RQ with elongation \\
 & on either side of the radio core \\
SDSS J232721.96+152437.3 & not in FIRST catalog; \\
 & NVSS shows extended weak lobes \\

\hline
\end{tabular}

\end{table*}

Since they are not classical RL sources some of the RI might be pre-
and/or post-cursors of the classical RL phenomenon. Support for this
interpretation might come from observations showing a flat or curved
radio spectra \citep{Falcke96a, O'Dea98}. Sources in this restricted
regime merit multifrequency radio measures. The most interesting RI
source in this context involves SDSS J232721.97+152437.3 (Table 3)
that is indicated with an ``X'' in Figure 5. It is RI based on an
integrated radio luminosity logL$_{1.4GHz}$=31.2 erg s$^{-1}$
Hz$^{-1}$. The NVSS radio map shows very extended weak lobes and a
strong core (NVSS core/lobe$\sim$3.4). It was not observed by FIRST
but it is unlikely that FIRST would have detected the very extended
lobes. The low surface brightness in these lobes may be the
signature of a past episode of RL activity with the bright core
possibly indicating a renewed phase of radio activity. We may be
observing this source between radio outbursts when old decaying
lobes can still be detected. A possibly related RL analog involves
SDSS J110538.99+020257.4. FIRST detects a strong core elongated
$\sim$45$^{o}$ to the direction of very low surface brightness
lobes, only one bright enough to be listed in the FIRST source
catalog. This is likely another example of a two-phase RL with very
old and very young lobe structures. Without doubt a few RI and/or RQ
involve weak lobes but there is no evidence for a large population
that could boost many CD sources in the range
logL$_{1.4GHz}$=31.0-31.6 erg s$^{-1}$ Hz$^{-1}$. Or one could
extend this range to include all of the so-called RL CDs fainter
than logL$_{1.4GHz}$$\sim$32.5. Deeper maps (e.g. PG 1309+355 in
\citealt{Falcke96b}; \citealt{Ulvestad05}) have failed to turn up
weak lobe structures.

The RL boundary is also similar to the FRI/FRII break at $\sim$
logL$_{1.4GHz}$=32 erg s$^{-1}$ Hz$^{-1}$ \citep{OL89}. It is
important to point out that the FRI/FRII break is not a sharp one
and may be a function of optical as well as radio luminosity
\citep{OW91, LO96}. The recent deep radio survey of RQ quasars
\citep{Leipski06} revealed several RQs with elongated, complex and
even double sided structures. One source (out of 14) shows (VLA
B-array) structure (PG0026+129) reminiscent of FRI morphology but on
a very small ($\sim$1.5 kpc) scale and is 1.5 orders of magnitude
less radio luminous than the weakest FRII in our RL sample leading
us to conclude that this level of activity is unrelated to the
classical RL activity. FRI sources are essentially absent from our
type 1 QSO sample. We found no FRI structures on FIRST and NVSS maps
for any of our RL/RI/RQ sources (Table 3).

\subsection{Biases?}

Is there a chance that we missed some RI, or especially RL QSOs, in
Population A? These could  expand the RL domain in Figure 4
lessening the strong RQ-RL difference that we found. As pointed out
before we should have detected all RI and RL sources in our
optically selected SDSS samples. We consider four possible sources
of bias in this study.

a) It is known that SDSS is biased against very narrow broad
emission line QSOs (often called Narrow Line Seyfert 1's=NLSy1s)
because at least one line with FWHM$>$1000 kms$^{-1}$ is required to
be assigned QSO type. We attempted to identify all such extreme
NLSy1 sources within our redshift and magnitude limits that are
assigned ``Galaxy'' type by the SDSS spectroscopic pipeline
\footnote{For such a task we used the SQL (Structured Query
Language), instead of a direct selection through the Spectroscopic
Query Form of SDSS; in the Appendix we reproduce the ``where''
clause we formulated.}. We found 97 sources with psf g $<$ 17.5 or
psf i $<$ 17.5 (see Table 2) and only one is clearly RL (SDSS
J150324.77+475829.6, but it shows a spectrum with a continuum in
high phase and almost missing H$\beta$) and one is RI (SDSS
J163323.58+471858.9). One of these sources actually shows the
broadest known FWHM H$\beta$ ($\sim$40000 km s$^{-1}$,
\citealt{Wang05}) which exceeded the comprehension of the SDSS broad
line identifier. These population A sources clearly show a small
probability of radio-loudness. Eventual addition of such NLSy1
sources (our template could not reduce them properly) would increase
the RQ-RL domain occupation difference discussed in \S~4.

b) SDSS is also apparently biased against steep-spectrum
lobe-dominated quasars \citep{Richards02}. There is no reason to
assume that any missed FRII would preferentially populate region A
in 4DE1. We tried to avoid missing RL sources with largely separated
radio lobes and no radio detected core between them by carefully
examining FIRST and NVSS radio maps. We consider that our approach
is very effective in turning up all FRIIs without a detected
radio-core at/near the position of quasar.

c) We must also consider that SDSS does not include sources brighter
than i $\sim$ 15.0. This is a technical limitation imposed to avoid
contamination of adjacent fibers by very bright sources. How would
the omission of these bright (mostly low luminosity Seyferts) AGN
affect our conclusions?  The latest incarnation of the 4DE1
spectroscopic Atlas \citep{Marziani03a} includes 215 objects of
which 61 are brighter than V=15.0 (4 RL) and 21 are brighter than V
=14.0 (3 RL). 56/61 objects show z $<$ 0.1 (41/61 with z $<$ 0.05).
Most of these objects (53) show bolometric luminosity
logL$_{bol}$$>$44.0 erg s$^{-1}$, which according to Figure 6 are
bright enough to be RL. The RL percentage ($\sim$7\%) suggests that
SDSS exclusion of such bright AGN will not affect our conclusions.

d) As mentioned earlier we have a total sample size of n=1770 QSOs
brighter than psf g=17.5 or psf i=17.5. We have almost complete
spectroscopic coverage for all RL and RI sources in this range
meaning that we extracted reliable 4DE1 parameters for use in
Figures 4 and 5. Similar parameters were extracted for n=333 psf g
selected RQ quasars. Would inclusion of fainter psf g selected and
the many psf i selected RQ sources change our 4DE1 definition/domain
of RQ? We think the answer is clearly ``no'' for several reasons: 1)
a random sample of 333 RQ is sufficient to define the general
properties of the RQ parent population, 2) the RQ domain defined
with the SDSS sample is very similar to that defined from our Atlas
sample \citep{Marziani03a, Sulentic07} that shows only partial
overlap with SDSS, 3) recent VLT spectroscopy of the H$\beta$ region
in high z sources \citep{Sulentic04, Sulentic06, Marziani08} again
show the same trends as for the low redshift samples. The psf i
selected quasars (fainter than psf g = 17.5) show too low S/N to
allow accurate 4DE1 measures to be extracted. They are in addition
strongly host galaxy contaminated as a class. Random examination of
these noisy spectra give no evidence that they would change the
general RQ properties derived from the brighter sources.

\subsection{Black Hole Mass and Radio Loudness}

The results of Figure 4 indicate that the vast majority of RL
sources show FWHM H$\beta$ $>$ 4000 km s$^{-1}$, with the median
FWHM H$\beta$ of the FRII population (viewed as inclined RL sources)
near 6750 km s$^{-1}$. Using FWHM H$\beta$ and L$_{5100\AA}$
measures to estimate black hole masses and L$_{bol}$/L$_{Edd}$
values implies that M$_{BH}$$>$1$\times$10$^8$ M$_{\odot}$ for RL
sources, with a strong concentration between 3$\times$10$^8$ and
3$\times$10$^9$ M$_{\odot}$ \citep{Marziani03b, Sulentic06}.
Eddington ratios for RL sources are restricted to
L$_{bol}$/L$_{Edd}$ $<$ 0.15-0.30 (see previous discussion). The RQ
majority show generally smaller values of M$_{BH}$ and larger values
of L$_{bol}$/L$_{Edd}$ \citep[e.g.][]{Boroson02, Marziani03b,
Dunlop03, MJ04, MM06}. However, some studies (e.g.
\citealt{Kording06b}) suggest that radio-loudness is not directly
related to a single variable like M$_{BH}$ or accretion rate.

RL/RQ comparisons using spectral properties are sensitive to the
relative contributions of Population A/B RQ sources in the sample
under study. RQ Population B sources will show masses and Eddington
ratios similar to RL sources while the majority of RQ sources
(Population A) will not. The numbers quoted here come from analysis
of our Atlas sample \citep{Marziani03a, Sulentic06}. Preliminary
analysis of the SDSS sample (e.g. Figure 4) indicate that the
conclusions will be very similar to those summarized here (see also
\citealt{Laor03} for some relevant comments).

A full discussion and comparison with other studies will be given in
a later paper \citep{Zamfir08}. We emphasize the importance of using
adequate S/N spectra and proper identification/subtraction of narrow
H$\beta$ for estimating black hole masses and Eddington ratios in
AGN.

\section{Conclusions}

Three criteria have been used to isolate RL quasar samples from the
RQ majority: 1) radio/optical flux density ratios (e.g. R$_{K}$ as
defined in \S~1), 2) radio luminosity and 3) radio morphology. The
first criterion is the least precise and was not used in this study.
We used a combination of criteria 2 and 3 which involved determining
the cutoff from the radio luminosities of the weakest FRII sources
using one of the most complete RL samples ever compiled. The cutoff
value log L$_{1.4GHz}$=31.6 ergs $s^{-1}$ Hz$^{-1}$ agrees closely
with the value derived in an earlier attempt using a more
heterogeneous sample. We think this value is therefore a robust
boundary for the classical RL phenomenon. We find many CD RL sources
that are RL by this definition but are on the low luminosity side of
most FRII sources. These RL CD sources cannot be FRII sources viewed
jet-on and are either boosted RQ quasars or precursors of the RL
phenomenon.

We find that RQ-RL comparisons involving  radio and bolometric
luminosity (diagrams) yield ambiguous results about the reality of a
RQ-RL dichotomy. A gap or dichotomy between the two populations is
filled by the radio bright end of the RQ source distribution and
possible radio pre- and/or post-cursors in the zone of RI (and RL)
sources. The optical diagnostic plane of 4DE1 provides much less
ambiguous evidence that RL show significant structural and kinematic
differences from the majority of RQ sources which is consistent with
a real dichotomy. 4DE1 also shows that RI and RQ sources are
spectroscopically indistinguishable. Our Type 1 QSO sample shows no
evidence for an FRI population within the radio resolution
constraints of NVSS/FIRST.

\section*{Acknowledgments}

S. Zamfir acknowledges support from the Graduate Council
Research/Creative Activity Fellowship offered by the Graduate School
of the University of Alabama for the 2007/2008 academic year. S.
Zamfir and J. W. Sulentic kindly acknowledge the hospitality of
Asiago Observatory (Italy) in the summer of 2006, when this project
was outlined. The authors thank the reviewer, Dr. Sebastian Jester,
for numerous comments and suggestions, which certainly improved the
quality of the paper.

Funding for the SDSS and SDSS-II has been provided by the Alfred P.
Sloan Foundation, the Participating Institutions, the National
Science Foundation, the U.S. Department of Energy, the National
Aeronautics and Space Administration, the Japanese Monbukagakusho,
the Max Planck Society, and the Higher Education Funding Council for
England. The SDSS Web Site is http://www.sdss.org/. The SDSS is
managed by the Astrophysical Research Consortium for the
Participating Institutions. The Participating Institutions are the
American Museum of Natural History, Astrophysical Institute Potsdam,
University of Basel, University of Cambridge, Case Western Reserve
University, University of Chicago, Drexel University, Fermilab, the
Institute for Advanced Study, the Japan Participation Group, Johns
Hopkins University, the Joint Institute for Nuclear Astrophysics,
the Kavli Institute for Particle Astrophysics and Cosmology, the
Korean Scientist Group, the Chinese Academy of Sciences (LAMOST),
Los Alamos National Laboratory, the Max-Planck-Institute for
Astronomy (MPIA), the Max-Planck-Institute for Astrophysics (MPA),
New Mexico State University, Ohio State University, University of
Pittsburgh, University of Portsmouth, Princeton University, the
United States Naval Observatory, and the University of Washington.

This research has made use of the NASA/IPAC Extragalactic Database
(NED) which is operated by the Jet Propulsion Laboratory, California
Institute of Technology, under contract with the National
Aeronautics and Space Administration.

\appendix
\section{An SQL Search for QSO with FWHM H$\beta$ $<$ 1000 km s$^{-1}$ in DR5}

We reproduce here the ``where'' clause we formulate to isolate
objects labeled ``Galaxy'' by the spectroscopic algorithm of SDSS
(DR5); their spectra show H$\beta$ emission line narrower than 1000
km s$^{-1}$ in the sources rest frame. The $\sigma$ interval
required in the query would translate into an \textit{observed} FWHM
H$\beta$ range $\sim$ 300-1700 km s$^{-1}$. This takes into account
the (1+z) scaling of line width from source's frame to the observed
frame, e.g. a rest frame FWHM H$\beta$ = 1000 km s$^{-1}$ would be
observed from z=0.7 as 1700 km s$^{-1}$.

\begin{verbatim}
SELECT ... FROM SpecPhoto as S,
     SpecLine as L
WHERE
      S.SpecObjID = L.SpecObjID and
      L.LineId = 4863 and
      S.z <= 0.7 and
      L.ew > 0 and
      L.sigma <= 11.7 and
      L.sigma >= 2.1 and
      S.psfMag_g <= 17.5
\end{verbatim}

The spectra selected this way have been visually examined and we
kept only the bona-fide Type 1 QSO for our statistical estimates, as
explained in the text. We selected ``Galaxy'' spectra based on psf i
magnitude cut replacing the last line in the ``where'' clause above
with: \begin{verbatim} S.psfMag_i <= 17.5 \end{verbatim}

\label{lastpage}
\end{document}